# Synthesis, structure and magnetization studies on $(Fe_{1-x}Cr_x)_2TeO_6$ tellurate system

K. D. Singh Mudher[1*], K. Krishnan[1] and I. K. Gopalakrishnan[2]

*[1]Fuel Chemistry Division, [2]Novel Materials and Structural Chemistry Division*
*Bhabha Atomic Research Centre, Mumbai 400085, India*

Darshan C. Kundaliya

*Center for Superconductivity Research, Department of Physics, University of Maryland, College Park, MD-20742, U.S.A*

and

S.K. Malik*

*Tata Institute of Fundamental Research, Colaba, Mumbai 400005, India*

## Abstract

The effect of substitution of Fe by Cr on the structural and magnetic properties of $Fe_2TeO_6$ in the system $(Fe_{1-x}Cr_x)_2TeO_6$, $(0.0 \leq x \leq 1.0)$ has been investigated on polycrystalline samples synthesized by solid state reaction. The lattice parameters are found to obey Vegard`s law in the entire composition range. The Néel temperature decreases linearly with increasing Cr content from ~230 K for x=0 to 90 K for x=1 and is attributed to the weakening of antiferromagnetic exchange interaction as $Cr^{3+}$ replaces $Fe^{3+}$ in $Fe_2TeO_6$.

*Corresponding authors: S.K. Malik Email: skm@tifr.res.in
Fax: +91-22-2280 4610; Tel: +91-22-2280 4545
K.D. Singh Email: kdsingh@apsara.barc.ernet.in

## Introduction

The compounds $Fe_2TeO_6$ and $Cr_2TeO_6$ are members of isomorphous series of compounds possessing an ordered inverse-trirutile structure with tetragonal symmetry (space group $P4_2/mnm$, No. 136) [1-3]. The structure can be regarded as a superlattice of rutile with a regular distribution of $Cr^{3+}$ or $Fe^{3+}$ and $Te^{6+}$ in octahedral coordination. Neutron diffraction studies by Kunnmann et al [4] have shown that these compounds are antiferromagnetically ordered at 4.2 K. The ordered state in $Fe_2TeO_6$ and $Cr_2TeO_6$ shows that the magnetic ions have collinear spin arrangements in a magnetic cell identical with the chemical cell with 3*d* ion moments of $4.19\mu_B$ and $2.45\mu_B$, respectively [4]. In $Fe_2TeO_6$, the spins are directed along the *c*-axis while in $Cr_2TeO_6$ these are directed along the basal plane. From magnetization and specific heat measurement, Yamaguchi and. Ishikawa [5] have shown that $Fe_2TeO_6$ and $Cr_2TeO_6$ have Néel temperatures ($T_N$) of 80 K and 230 K, respectively. In this paper we report the effect of substituting Fe by Cr in $(Fe_{1-x}Cr_x)_2TeO_6$ on the structural and magnetic properties in the entire composition range ($0.0 \leq x \leq 1.0$)

## Experimental

The compounds in the series $(Fe_{1-x}Cr_x)_2TeO_6$ were prepared by solid state reaction techniques from high purity $Fe_2O_3$ (99.9%), $Cr_2O_3$(99.9%) and $TeO_2$(99.99%). The components were weighed to obtain desired concentration (x = 0.0, 0.2, 0.4, 0.6, 0.8 and 1.0), mixed intimately and heated in alumina boats at 700 ºC for 20 h. After this initial firing, the standard procedure of repeated grinding and firing at 700 ºC was employed until X-ray diffraction analysis showed them to be single phase compounds The identification of the phases and determination of the lattice parameters was done by X-ray powder diffraction analysis using graphite monochromatatised CuKα radiation ($\lambda$=0.15406 nm) on a STOE diffractometer. Silicon powder was used as an external standard. Structural analysis was carried by Rietveld method for the X-ray intensity data collected for the range 10°∠2θ∠100°, counting for 5s at each step of 0.02°.

Magnetization measurements on $(Fe_{1-x}Cr_x)_2TeO_6$ compounds were carried out at various temperatures and fields using a SQUID magnetometer (MPMS, Quantum Design). The temperature dependence of the magnetization was measured in zero-field-cooled (ZFC) mode in the temperature range between 4.5 and 300 K. The field dependence of the magnetization was measured at different temperatures in magnetic fields upto 55 kOe.

## Results and Discussions

### *a. Structural studies:*

The compounds $Fe_2TeO_6$ and $Cr_2TeO_6$ crystallize in the tetragonal trirutile structure [4-6]. Powder X-ray diffraction (XRD) analysis of $(Fe_{1-x}Cr_x)_2TeO_6$, (0 0∠x∠1.0) samples show that the compounds with in-between compositions have the same crystal structure as the end members, thus forming a solid solution in the entire range of composition. This is evident from the data presented in Figures 1 which show the lattice parameters and molar volume (Inset figure) decrease linearly as a function of Cr content obeying Vegard`s law. The decrease can be attributed to the fact that ionic radius of $Cr^{3+}$ (0.62 Å) is smaller than that of $Fe^{3+}$ (0.65 Å) for 6-coordination. Figure 2 shows the detailed structural analysis performed on the XRD data of $(Fe_{1-x}Cr_x)_2TeO_6$, by Rietveld pofile refinement method with the computer code DBWS 11 [7], using atomic coordinates of $Cr_2TeO_6$. The occupancy factors for the Fe and Cr were refined along with other variables like scale factor, background parameters, halfwidth parameters defining pseudo-Voigt peak shape, the unit cell dimensions, etc. The structural parameters including atomic coordinates and isotropic thermal parameters of atoms and selected inter-atomic distances for $(Fe_{1-x}Cr_x)_2TeO_6$ are given in Table 1. In the trirutile structure, each Cr or Fe atom is surrounded by six oxygen atoms in an octahedral coordination. The cation oxygen octahedra form edge-sharing chains which are alternately occupied by Cr (or Fe)$O_6$ and $TeO_6$ octahedra in the ratio of 1:2.

***b. Magnetization studies***

Figure 3 (a) shows the plot of magnetization (ZFC) vs. temperature for the polycrystalline samples in the $(Fe_{1-x}Cr_x)_2TeO_6$ system with x = 0, 0.1 .15, .25, 5, .75, .9 and 1.0. The isothermal magnetization data of the end compounds, $Fe_2TeO_6$ and $Cr_2TeO_6$, as a function of field at different temperatures are presented in Figures 3(b) & (c), respectively. It can be seen from Figure 4 that the magnetization of the samples passes through a broad maxima as the temperature is lowered. The end compounds $Fe_2TeO_6$ and $Cr_2TeO_6$ being antiferromagnets [4-6], it is reasonable to attribute the magnetization maxima observed in the samples in $(Fe_{1-x}Cr_x)_2TeO_6$ system to the Néel temperature ($T_N$). This is also evident from the isothermal magnetiztion data of end compounds presented in Figs. 3(b) & (c). It can be seen that magnetization is linear and without any hysteresis. The variation of $T_N$ in $(Fe_{1-x}Cr_x)_2TeO_6$ with concentration x is plotted in Fig. 4. It can be seen that $T_N$ decreases as the Cr concentration increases. The decrease of $T_N$ is linear upto x = 0.75, after which $T_N$ remains nearly constant at 90 K. The Néel temperatures of the end compounds are in good agreement with those reported in the literature [5].

Figure 5 depicts the trirule structure of $Fe_2TeO_6$. Small filled circles with arrows indicate Fe atoms while small open circles indicate Te atoms. Oxygen is represented by large open circles. The arrows indicate the directions of the $Fe^{3+}$ ion magnetic moment. Antiferromagnetism in $Fe_2TeO_6$ has been attributed to the exchange interaction between $Fe^{3+}$ ions through Fe-O-Fe pathway [4, 6]. The exchange interaction between $Fe^{3+}$-ions in $Fe_2TeO_6$ takes place through an intervening $O^{2-}$ ion and is referred as supercxchange interaction [8]. In $Fe_2TeO_6$, there are two different Fe-O-Fe pathways for antiferromagnetic superexchange interaction as depicted in Fig. 5. The first one is the antiferromagnetic ordering of Fe layers into double layers without correlation of these double layers with each other. The ordering of two neigbouring Fe layers would be accomplished by antiferromagnetic superexchange between next-nearest- neigbour (nnn) $Fe^{3+}$ ions through an oxygen ion with an Fe-O-Fe angle of $125^o$ (marked by doted line in Figure 5). In this way, the magnetic ordering is restricted to the bilayer without correlation along the c-axis due to the intervening Te layer. The second pathway (marked by a dashed line) is the antiferromagnetic superexchange between nearest $Fe^{3+}$ ions at an angle of $100^o$. The two possible pathways available for superexhange antiferromagnetic interactions in $Fe_2TeO_6$ are not equivalent. This may explain the rather rounded maxima observed around $T_N$ in $Fe_2TeO_6$. $Cr_2TeO_6$ is structurally equivalent to $Fe_2TeO_6$ except that the spins are directed along the basal plane [4]. The observed decrease in the Néel temperature ($T_N$) with increasing Cr content in $(Fe_{1-x}Cr_x)_2TeO_6$ may be attributed to a weakening of antiferromagnetic exchange interaction as $Cr^{3+}(d^3)$ replaces $Fe^{3+}(d^5)$.

In conclusion, we have synthesized trirutiles in the $(Fe_{1-x}Cr_x)_2TeO_6$ system. X-ray diffraction analyses of the samples show that they form solid solution in the entire composition range ($0.0 \leq x \leq 1.0$). DC magnetization studies as a function field and temperature show that all these compounds are antiferromagnetic. The Néel temperature, $T_{N,}$ decreases from 230 K to 90 K as x increases from 0.0 to1.0.

**Figure Captions**

Figure 1. The variation of lattice parameter *a* and *c* of $(Fe_{1-x}Cr_x)_2TeO_6$ as a function of Cr content (x). Inset shows the variation of molar volume of $(Fe_{1-x}Cr_x)_2TeO_6$ as a function of Cr content (x)

Figure 2. The observed (dots) and profile fitted (curves) room temperature X-ray diffractogram of a typical $(Fe_{1-x}Cr_x)_2TeO_6$. The bottom trace is the difference pattern. The vertical bars indicate peak positions.

Figure 3. (a) Temperature dependence of magnetization (ZFC) of $(Fe_{1-x}Cr_x)_2TeO_6$ at an applied field of 100Oe, (b) Isothermal magnetization of $Fe_2TeO_6$ at different temperatures, (c) Isothermal magnetization of $Cr_2TeO_6$ at different temperatures.

Figure 4. The variation of Néel temperature ($T_N$) as a function of Cr content in $(Fe_{1-x}Cr_x)_2TeO_6$.

Figure 5. Trirutile magnetic structure of $Fe_2TeO_6$.

Table 1. Structural parameters for (Fe,Cr)$TeO_6$ as determined by Rietveld analysis of X-ray data

| | |
|---|---|
| Wavelength (Å) | 1.5406 |
| a (Å) | 4.5760(2) |
| b (Å) | 4.5760(2) |
| c (Å) | 9.0464(4) |
| Space group | $P4_2/mnm$ |
| $R_p$ (%) | 11.2 |
| $R_{wp}$ (%) | 12.7 |
| $R_{exp}$ (%) | 8.8 |
| Goodness of fit | 1.19 |

*Positional parameters*

| Atom | site | x | y | z | B($Å^2$) |
|---|---|---|---|---|---|
| Cr or Fe | 4e | 0 | 0 | 0.3341(4) | 1.5 |
| Te | 2a | 0 | 0 | 0 | 1.2 |
| O1 | 4f | 0.3063(18) | 0.3063(18) | 0 | 2.5 |
| O2 | 8j | 0.3078(11) | 0.3078(11) | 0.3358(10) | 2.5 |

***Distances (Å) in octahedral coordination***

| | |
|---|---|
| Te-O1×2 | 1.982(8) |
| Te-O1×4 | 1.937(7) |
| Cr/Fe-O1×2 | 1.955(6) |
| Cr/Fe-O1×2 | 1.977(8) |
| Cr/Fe-O1×2 | 1.992(5) |

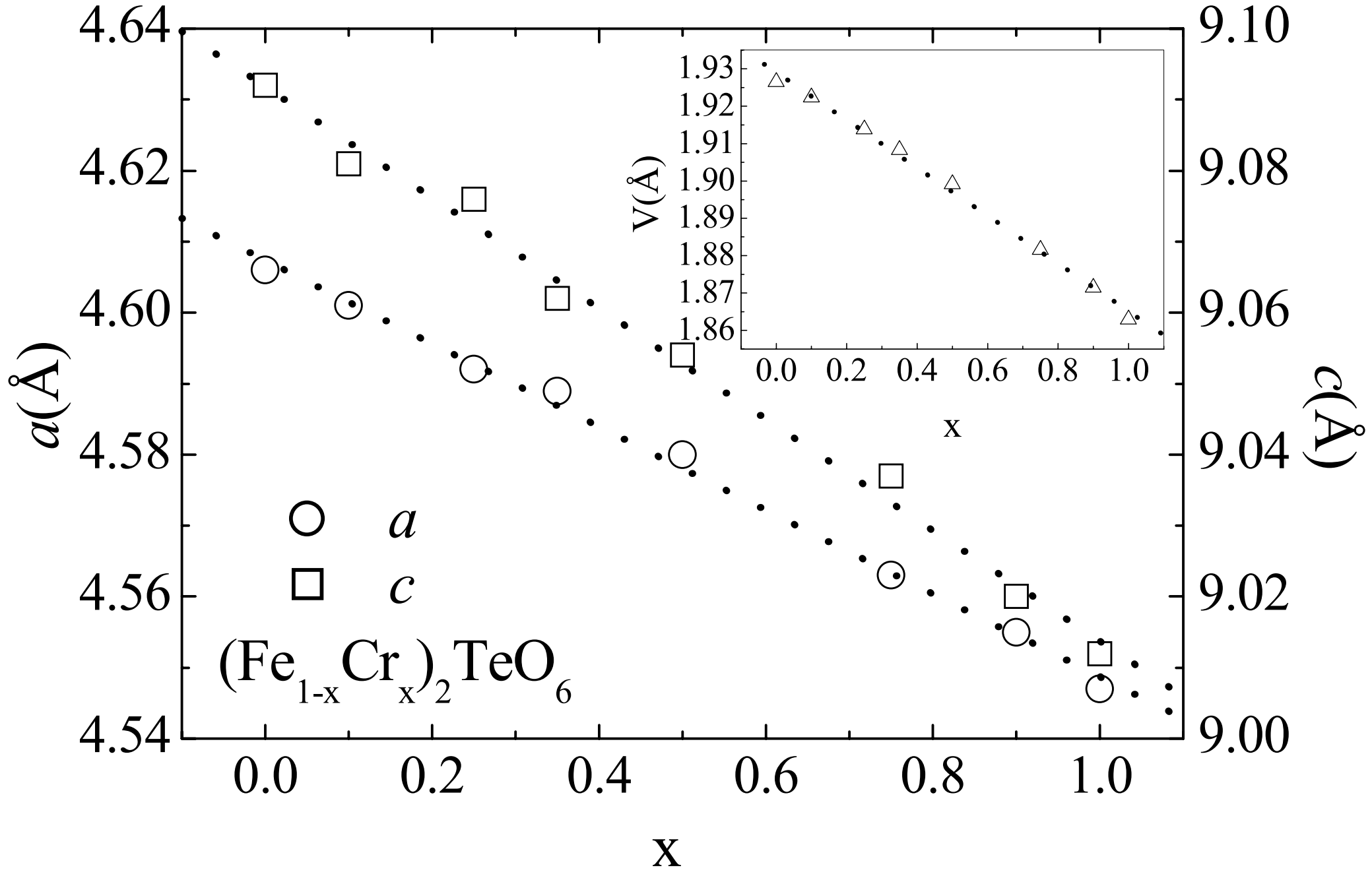

Figure 1 *Singh et al.*

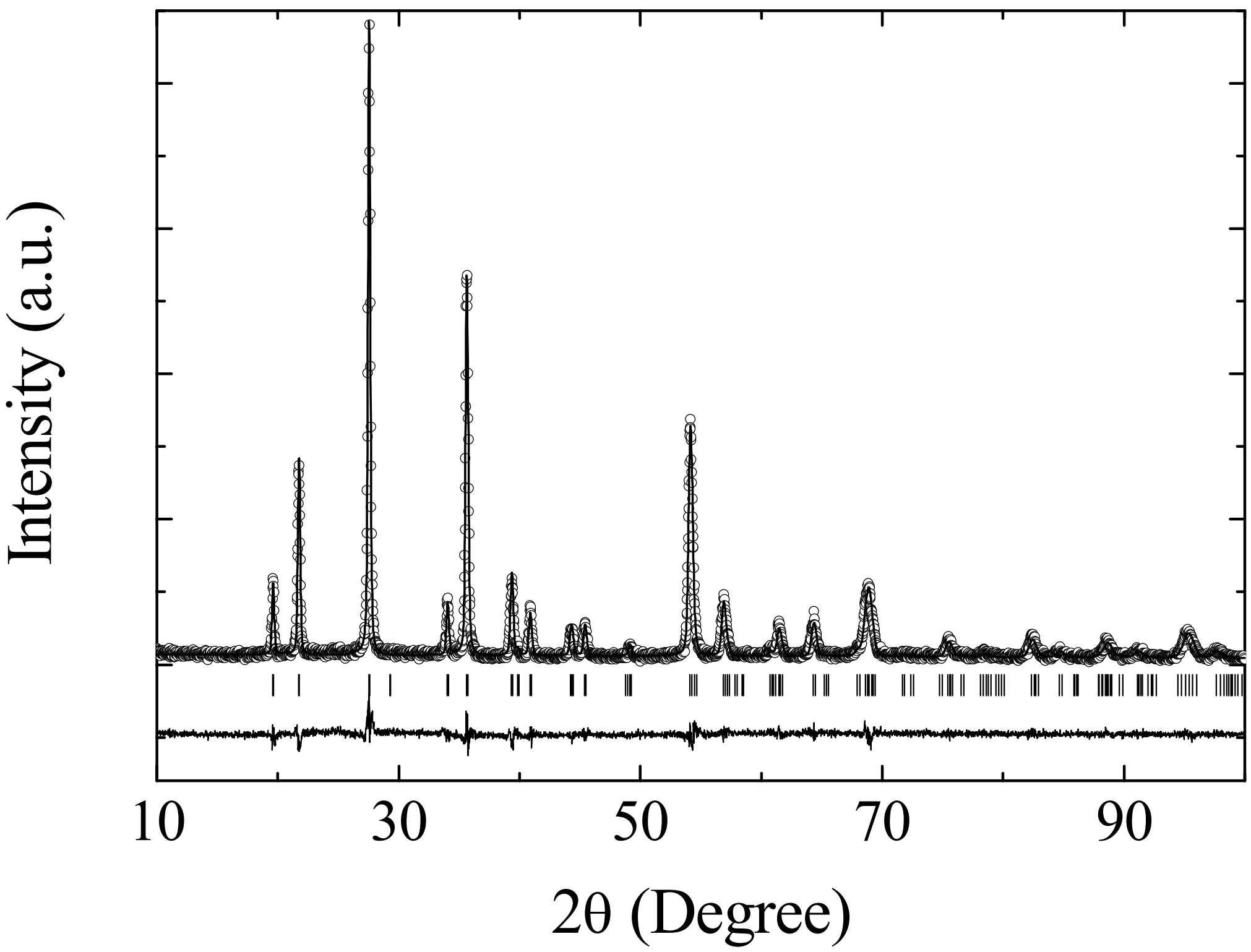

Figure 2 *Singh et al.*

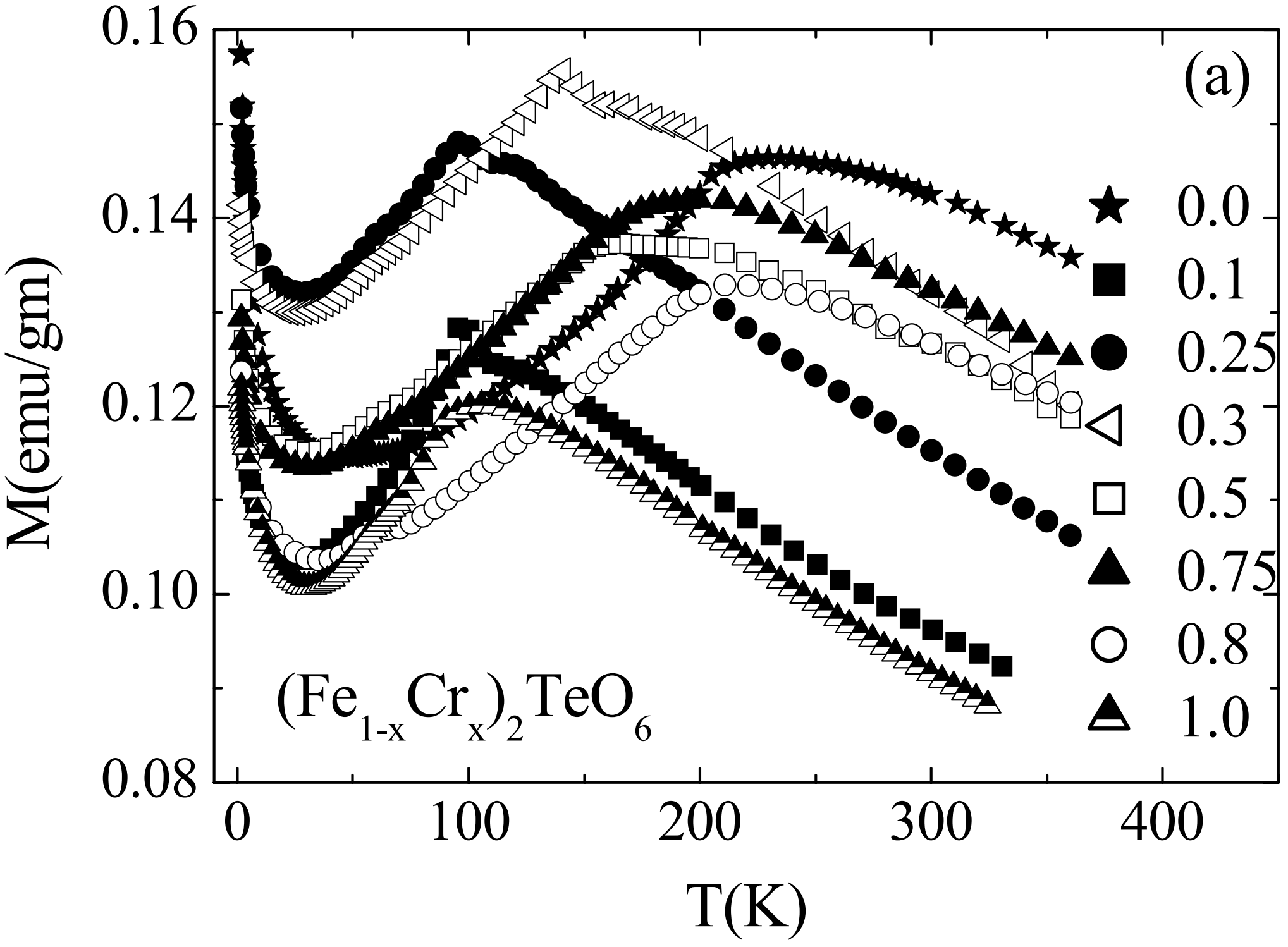

Figure 3 (a) *Singh et al.*

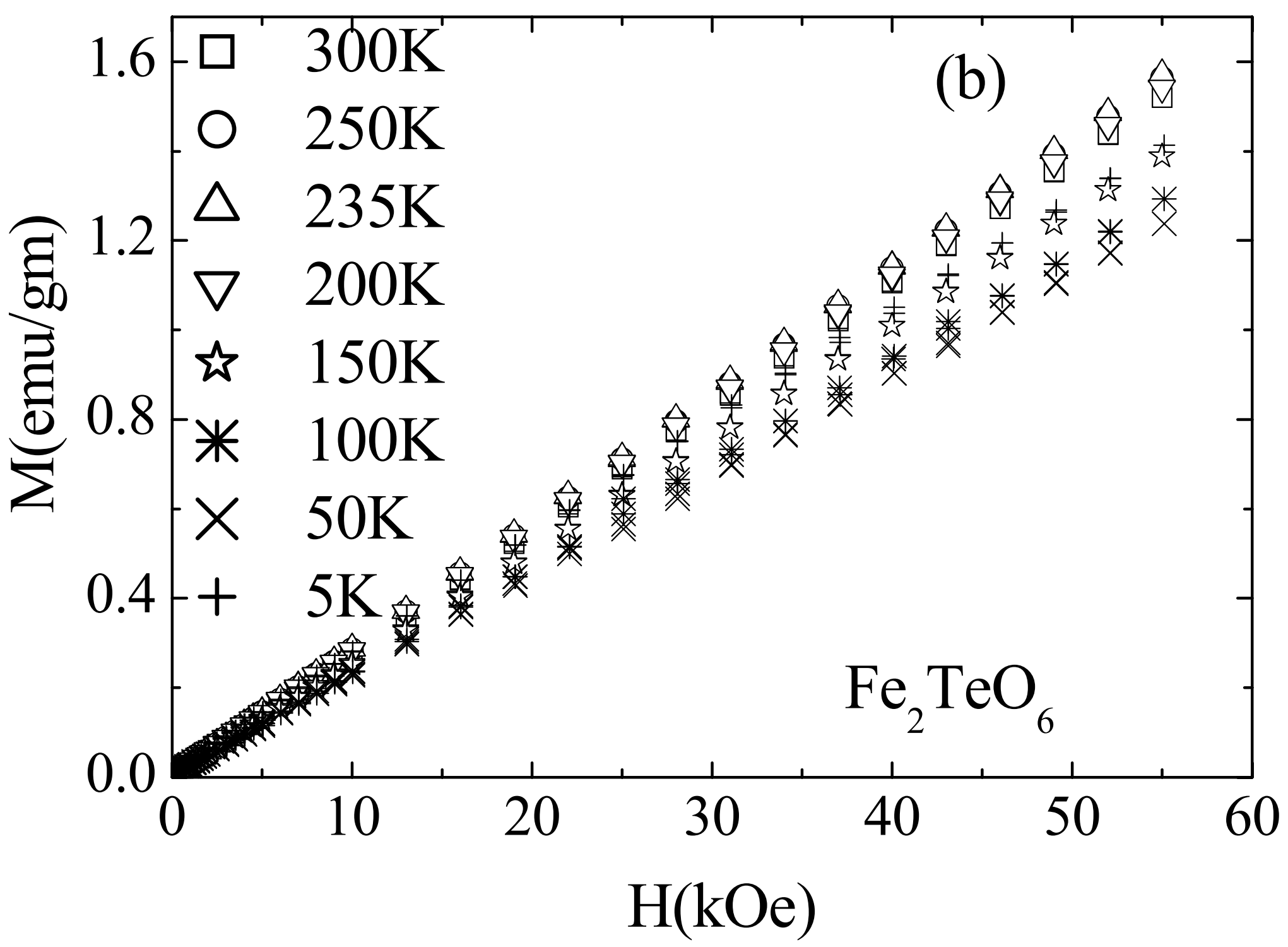

Figure 3 (b) *Singh et al.*

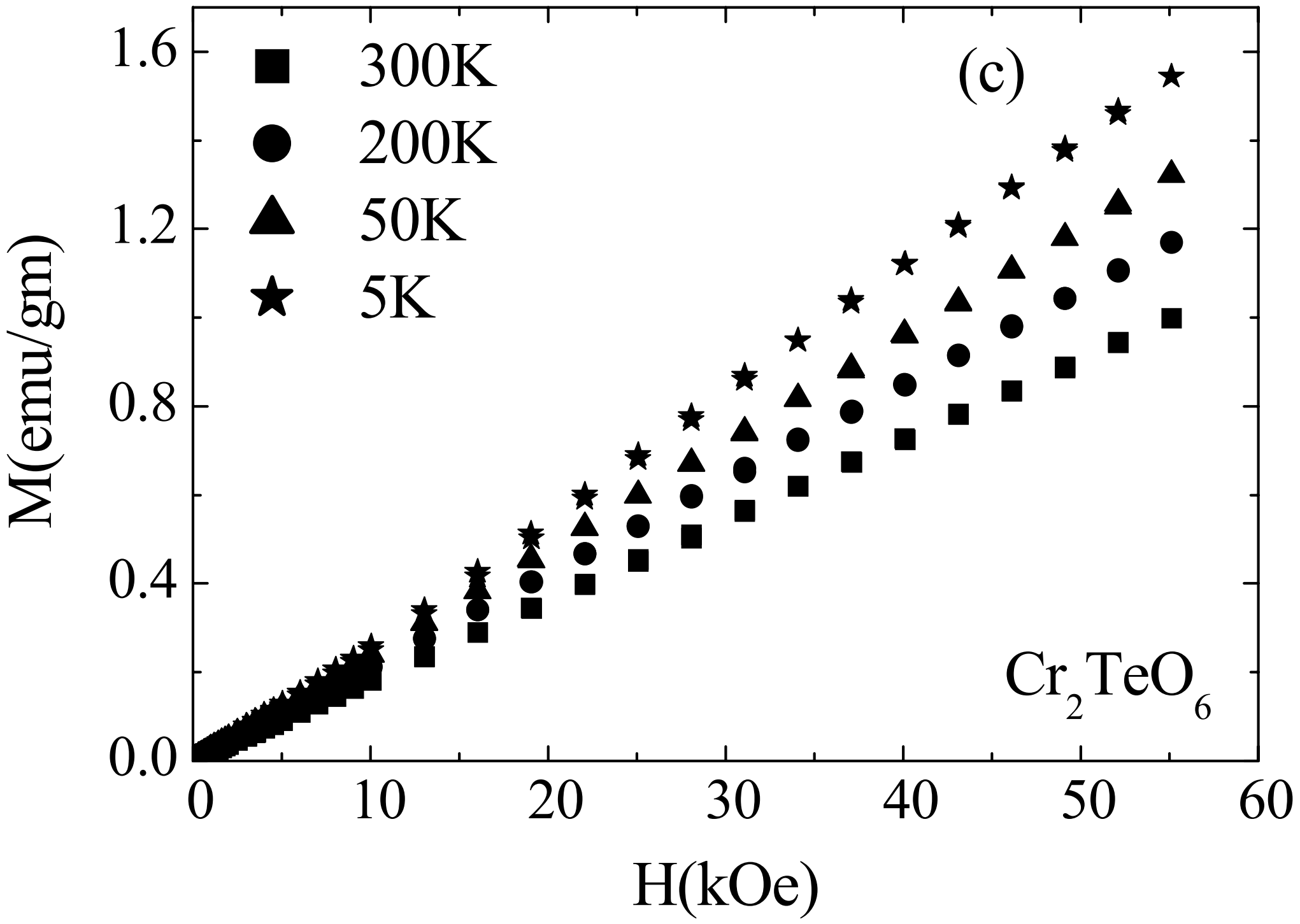

Figure 3(c) *Singh et al.*

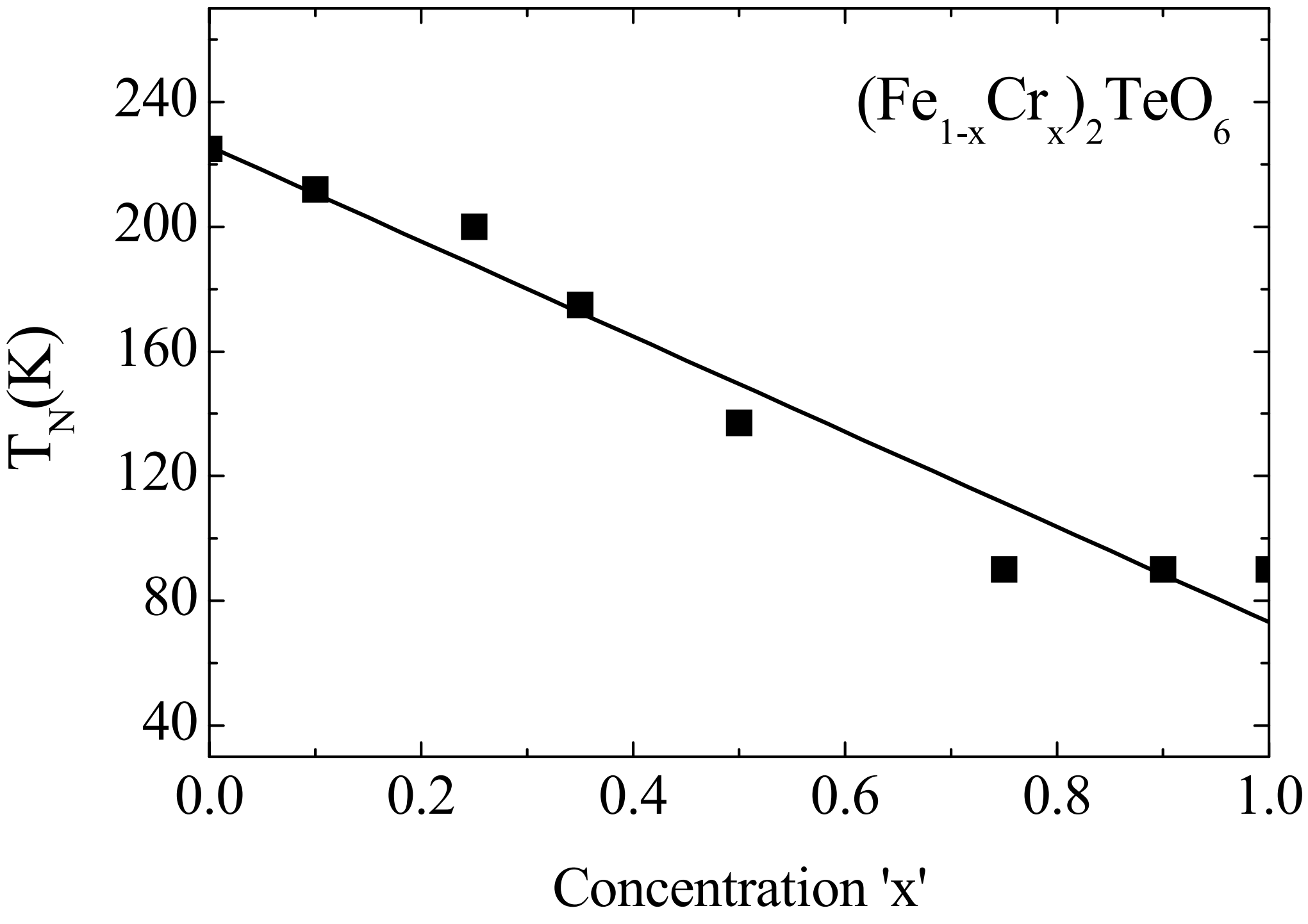

Figure 4 *Singh et al.*

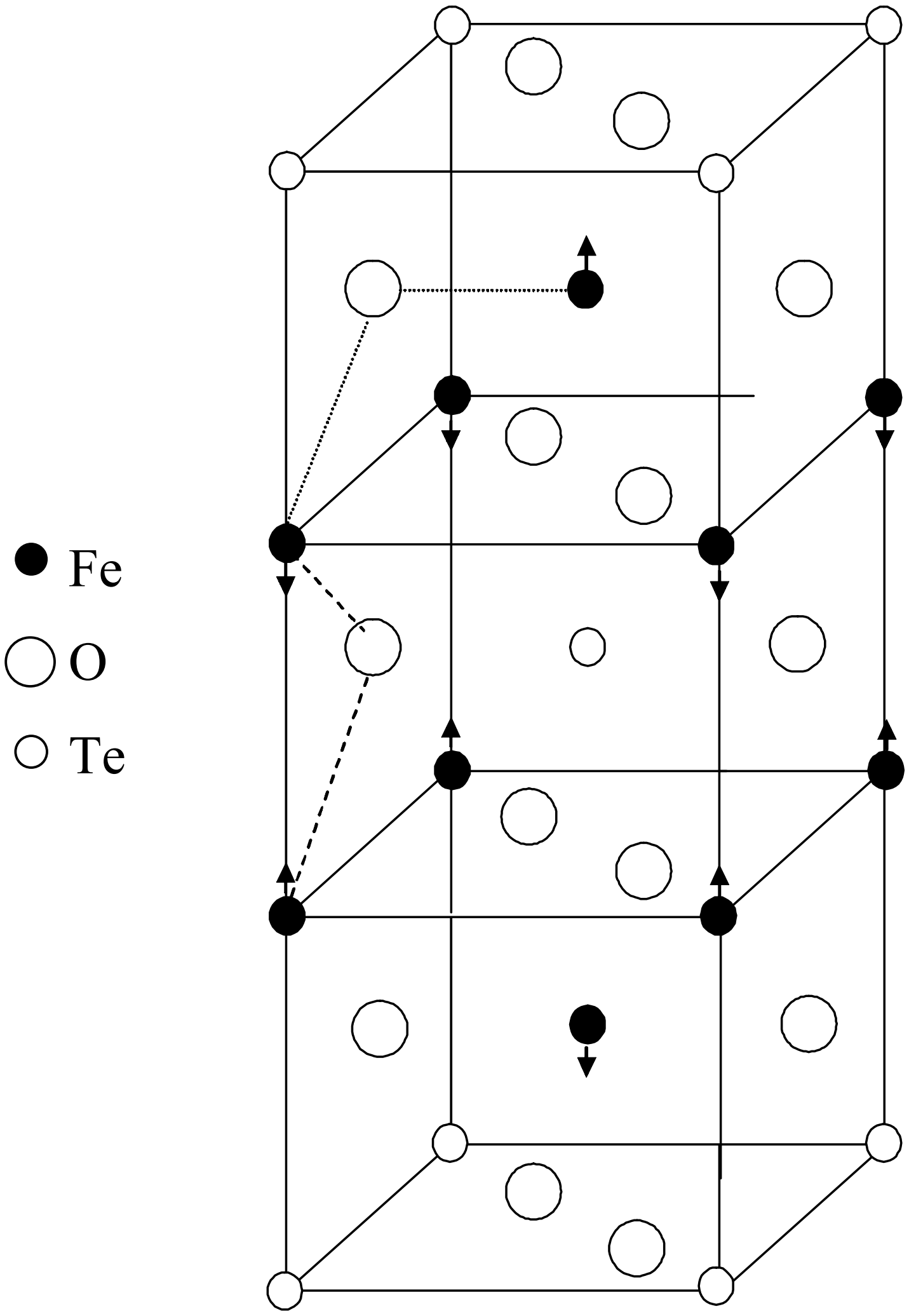

Figure 5 *Singh et al.*